\pdfoutput=1
\RequirePackage{fix-cm}
\documentclass[twocolumn]{svjour3}          
\smartqed  
\usepackage{graphicx}
\usepackage[round,authoryear,sectionbib]{natbib}
\begin{document}
\bibliographystyle{plainnat}

\title{A New Analysis Method for \\Simulations Using Node Categorizations
}


\author{Tomoyuki Yuasa         \and
        Susumu Shirayama 
}


\institute{T. Yuasa \at
              School of Engineering, the University of Tokyo\\
              Tel.: +81-3-5841-6552\\
              \email{yuasa@nakl.t.u-tokyo.ac.jp}           
           \and
           S. Shirayama \at
              School of Engineering, the University of Tokyo\\
              Tel.: +81-3-5841-6531\\
              \email{sirayama@sys.t.u-tokyo.ac.jp}
}

\date{Received: date / Accepted: date}

\maketitle

\begin{abstract}
Most research concerning the influence of network structure on phenomena taking place on the network focus on relationships between global statistics of the network structure and characteristic properties of those phenomena, even though local structure has a significant effect on the dynamics of some phenomena.
In the present paper, we propose a new analysis method for phenomena on networks based on a categorization of nodes.
First, local statistics such as the average path length and the clustering coefficient for a node are calculated and assigned to the respective node.
Then, the nodes are categorized using the self-organizing map (SOM) algorithm.
Characteristic properties of the phenomena of interest are visualized for each category of nodes. 
The validity of our method is demonstrated using the results of two simulation models. The proposed method is useful as a research tool to understand the behavior of networks, in particular, for the large-scale networks that existing visualization techniques cannot work well.
\end{abstract}

\section{Introduction}
\label{intro}
Many phenomena in the real world have been studied with respect to the network structure behind them \citep{boccaletti2006complex}. 
Numerical experiments with mathematical models to simulate such phenomena on networks are performed in most such studies. These networks are generated using some network model, in which each node and edge represents an agent and a relationship between the agents, respectively. The simulation proceeds by allowing the states of agents to evolve according to transition rules.

The analyses of such simulations using complex networks have revealed fundamental mechanisms of such phenomena as epidemic outbreaks \citep{pastor2001epidemicA,moreno2002epidemic,parshani2010epidemic}, decision making with respect to a social dilemma \citep{nowak2004nature,tomochi2004defectors,tsukamoto2010spd}, and synchronization of interactive units \citep{gomez2007paths}.
This type of analysis is also performed for some real networks such as the blogosphere\citep{cha2011spread}.

The analysis methods used in previous studies can mainly be classified into two types. One is to investigate the relationships between phenomena and the statistical properties of the network structure, such as the average path length and the clustering coefficient. The mechanisms of the phenomena are analyzed on the basis of the global structure of the network. This analysis is from a macroscopic standpoint.
The other type is based on the relationships between phenomena and the local characteristics of the nodes or edges, such as degree, node or edge betweenness, and the local clustering coefficient. This type of analysis explains the mechanisms of phenomena from a microscopic perspective based on the important nodes or edges in the network.
These two methods are often used simultaneously to examine the relationships between phenomena and the network structure. In most cases, it is obvious that these are insufficient to reveal the details of network phenomena, since the former methods lacks any local perspective, and using the latter methods, it is difficult to associate the role of important nodes with the global dynamics of phenomena. One of the most important issues for both types of method is how to connect the influence of local structure on phenomena with the global dynamics.

To address this issue, a visualization method in which the states of agents are visualized on the positions of nodes determined by a graph layout technique is sometimes used to analyze phenomena. Such visualization enables intuitive analysis using local and global structures of networks\citep{rosen2011social,adnan2011promoting,pham2011development}. However, the many possible layouts for the same network make interpreting the results difficult. In addition, as the number of nodes increases, the graph layout itself becomes more complicated \citep{van2008centrality,uchida2007formation}, making the extraction of useful information from a large-scale network visualization quite difficult.
Another method to address the abovementioned issue is to use the community structure of networks. The community structure is that which connects the local structure with the global one, and it can provide some mesoscopic perspective to the analysis \citep{newman2006finding,newman2006modularity,saravanan2011analyzing}.
However, since the community structure depends on the definition of the community and the extraction method, it is possible to extract different community structures from the same network\citep{fortunato2010community}. In this way, an analysis method based on community structure has a difficulty similar to that of visualization methods with respect to the interpretation of the results.

In the present paper, we propose a new analysis method for simulations using networks. Our method is based on the categorization of nodes. The nodes of the network being used in a simulation are categorized according to their local characteristics, and the simulation results for the network is visualized for each category of nodes. We apply our method to two simulations, and the validity of our method is discussed.

\section{Proposal Method}
\label{sec:1}
\subsection{Node Categorization}
\subsubsection{Characteristic properties of nodes}
Several of the statistical properties of network structures, average path length and the clustering coefficient, are obtained by averaging over the local values at each node. Therefore, such properties which can be calculated for individual nodes are used for our node categorization.

First, we define the property of  nodes as a multivariate variable ${\bf n}$. The variable ${\bf n}$ is composed of the degree, the average degree of neighboring nodes, the node betweenness, the average path length, and the clustering coefficient.

Let $N$ be the number of nodes in the network. The $i$-th node of the network is denoted by $v_i (i=1,..,N)$. 
The degree of the node $v_i$ is denoted by $k_i$.
The average degree of neighboring nodes of $v_i$ is denoted by $k^i_{nn}$ and defined as the average degree of nodes linked to $v_i$. 

The node betweenness of $v_i$ is denoted by $b_i$ and defined as the proportion of shortest paths between other pairs which include $v_i$. $b_i$ is calculated as follows.
Let $v_{i_s}$ and $v_{i_t}$ be the start and terminal nodes, respectively.
\begin{equation}
b_i = \frac{
\sum_{i_s=1;i_s\not= i} ^N
{
\sum_{i_t=1;i_t\not= i} ^{i_s-1}
}
\frac{g^{(i_s i_t)}_i}{N_{i_s i_t}}}{(N-1)(N-2)/2} 
\end{equation}
where $g^{(i_s i_t)}_i$ is the number of shortest paths between $v_{i_s}$ and $v_{i_t}$ via $v_i$, $N_{i_s i_t}$ is the total number of shortest paths between $v_{i_s}$ and $v_{i_t}$, and  the denominator is a normalization factor.

Let $L_i$ and $C_i$ be the average path length and clustering coefficient of $v_i$, respectively.
$L_i$ is calculated by
\begin{equation}
L_i = \sum_{i\not=j}\frac{d(v_i,v_j)}{N-1}
\end{equation}
where $d(v_i,v_j)$ is the length of the shortest path between $v_i$ and $v_j$. $C_i$ is calculated by
\begin{equation}
C_i = \frac{E_i}{k_i(k_i-1)/2}
\end{equation}
where $E_i$ is the total number of links existing between pairs of nodes adjacent to $v_i$.

The property of node $v_i$, multivariate variable ${\bf n}$, is ${\bf n}_i = (k_i, k_{nn}^i, b_i, L_i, C_i)$. ${\bf n}_i$ is calculated for each node and stored with the identifier of the network.

\subsubsection{Categorization method}
Systematical data mining operations are applied to the dataset which consists of ${\bf n}_i (i=1,...,N)$ in order to categorize the nodes.

\begin{figure}[t]
\begin{center}
\includegraphics[width=0.6\linewidth]{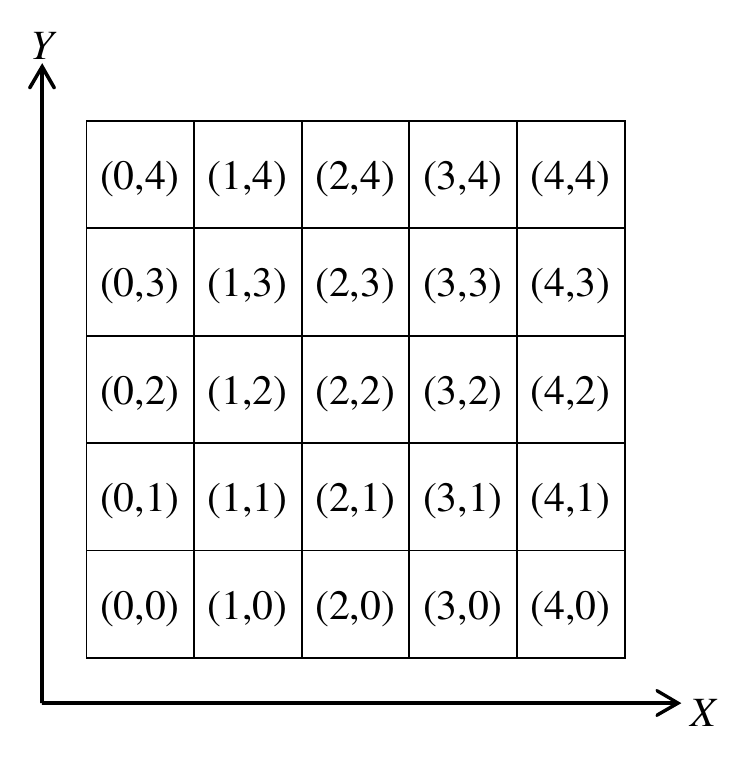}
\end{center}
\caption{Location of categories in the SOM}
\label{fig:lattice}
\end{figure}

In this paper, the nodes are categorized using the self-organizing map (SOM) algorithm.
The results from applying the SOM algorithm are displayed on a two-dimensional lattice. Figure \ref{fig:lattice} is an example of a 5 by 5 lattice. We consider each region to be a cell, which is identified as $(X,Y)$ according to the axes shown in Figure \ref{fig:lattice}. Let $M$ denote the cell. Using the SOM algorithm, the $N$ nodes are each assigned to one of the cells. 
Herein, each category corresponds to one of these cells (Figure \ref{fig:lattice}).

\begin{figure}[t]
\begin{center}
\includegraphics[width=1.0\linewidth]{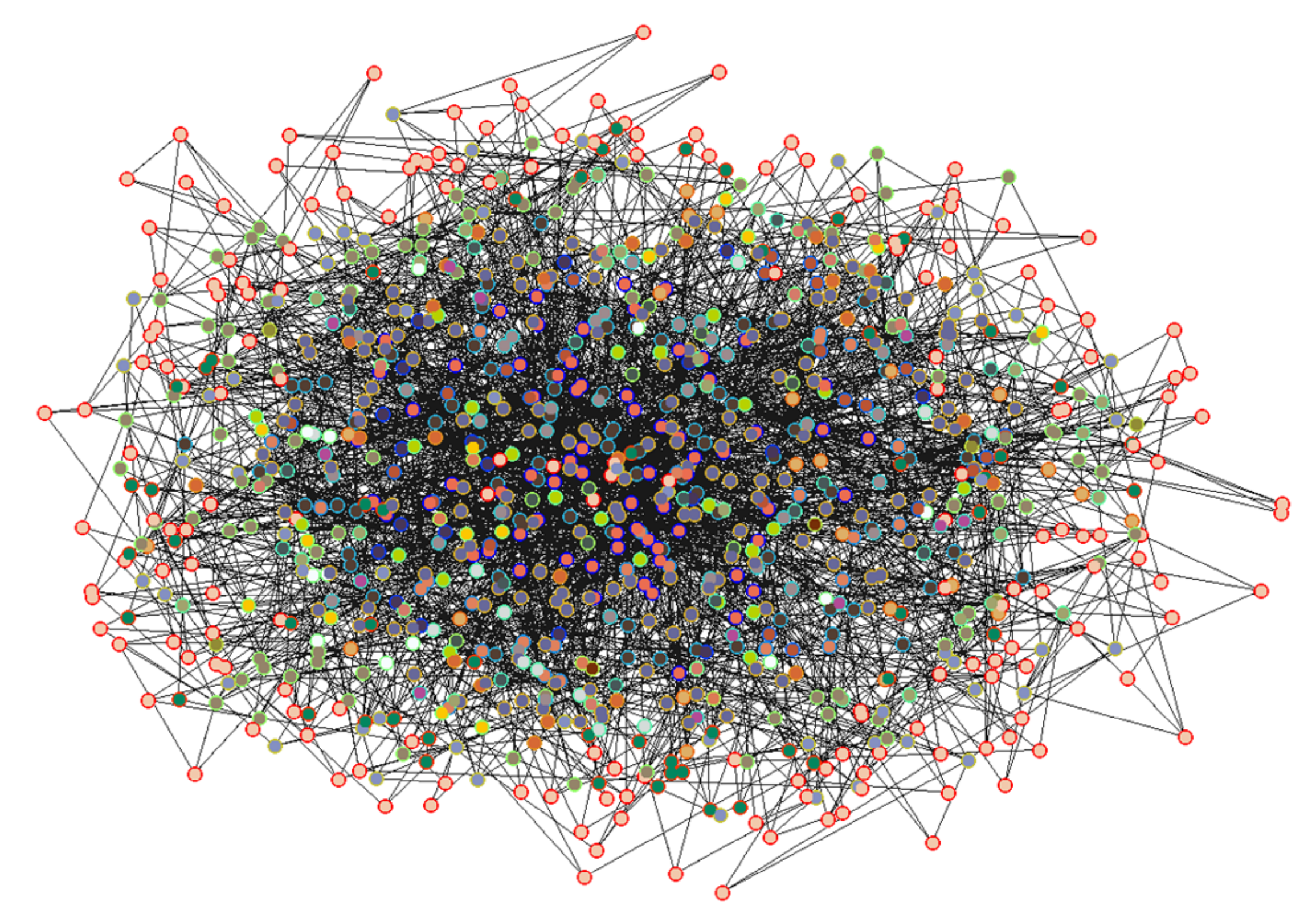}
\end{center}
\caption{Example of node categorization in a network}
\label{fig:ctg_ex}
\end{figure}

Figure \ref{fig:ctg_ex} shows an example of node categorization in a network.
The network is visualized by Pajek \citep{Pajek2003} using the Kamada-Kawai 
graph layout algorithm \citep{KK1989}. The nodes are colored corresponding to 
each category.

The results of categorization is checked using heat maps for each component of ${\bf n}$.
The heat maps are generated according to the cell average for each component, which are obtained for each cell by averaging over the nodes belonging to that cell.


\subsection{Visualization of Simulation Results}
In the present study, we are interested in simulations using networks. Each node and edge represents an agent and a relationship between the agents, respectively. The simulations proceed so that the state of an agent evolves by applying transition rules.

The nodes in the network used for the simulation are categorized according to their local characteristics. We visualize the simulation results on the network for each category, which contains agents with similar characteristics of the local network structure.

\begin{figure}[t]
\begin{center}
\includegraphics[width=1.0\linewidth]{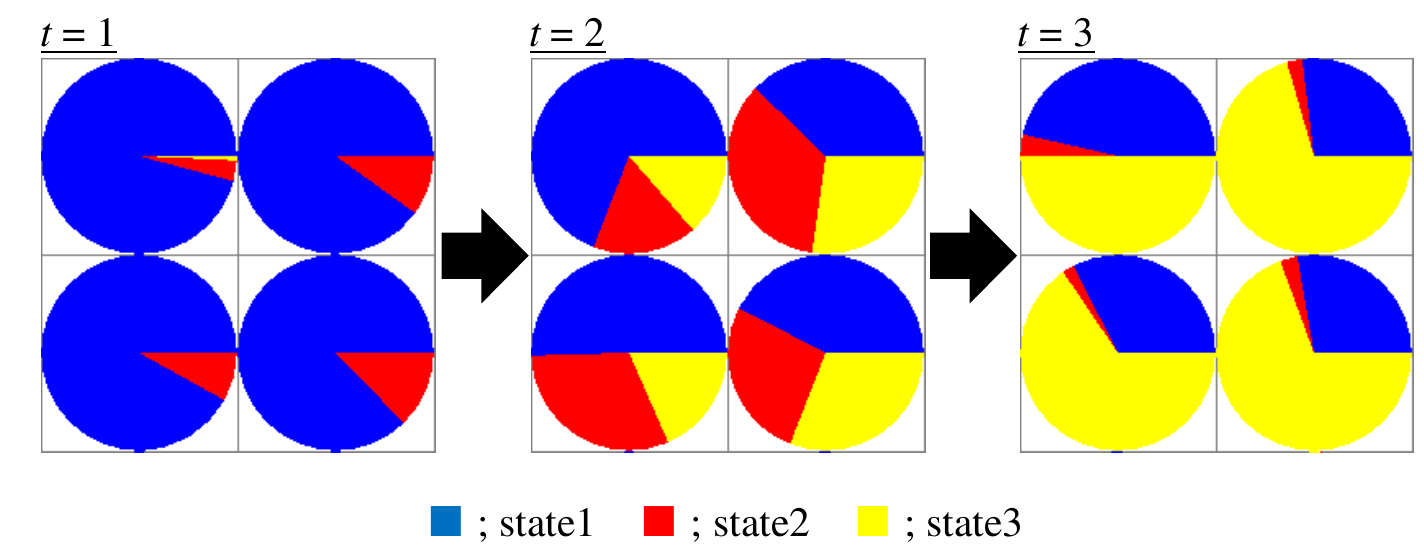}
\end{center}
\caption{Example of a visualization}
\label{fig:visual_ex}
\end{figure}

Herein, pie charts are used to visualize the results. The areas of the pie charts are proportional to the rate of each state of the agents in each category. The pie charts are displayed at the spatial location of each cell in the SOM. Time variation of the proportion in each category is also visualized using pie charts. Figure \ref{fig:visual_ex} is an example of this visualization.

Combining the heat maps which show the features of the categorized nodes with the pie charts which show the results of simulation, we analyze phenomena on the networks.

\section{Experiments and Results}
\label{sec:2}
\subsection{Simulation Models}
\subsubsection{Generation of networks}
In this paper, the networks used in the two simulations are generated using the HK model \citep{holme2002growing} and the CNN model \citep{vazquez2003growing}. Each network is composed of 10000 nodes ($N=10000$). The average degree $<k>$ is set to be about 8.

The HK model is that proposed by \cite{holme2002growing}. The network is generated by processes based on ``growth of the network", ``preferential attachment", and ``triad formation". The generated network has a scale-free property in which the degree distribution follows the power law. It has been shown that the clustering coefficient tends to be high in such networks.

The CNN model is that proposed by \cite{vazquez2003growing}. The network is generated by processes based on ``growth of the network" and ``change of potential links to real links". 
The generated network has a scale-free property. It has been shown that the clustering coefficient is high and the network becomes ``assortative", by which is meant that nodes with similar degrees tend to be linked to each other.

On these networks, both an epidemic propagation and a spatial prisoner's dilemma described below are examined.

\subsubsection{Epidemic propagation on networks}
We employ the SIR model for the epidemic propagation simulation. 
Each agent (node) takes one of three different states: S (susceptible or healthy), I (infectious), or R (removed, immunized, or dead). Each agent changes its state according to the states of neighboring agents.

Let $\lambda$ and $\mu$ be the infection and recovery rates, respectively.
The number of neighboring infectious agents is denoted by $n(\mathrm{I})$. 
At the beginning, all the agents (nodes) are in the state S.
Then, several agents are chosen randomly, and their states are changed to I.
The simulation proceeds by the following process after each time increment $dt$, where $dt$ is small:
\begin{itemize}
\item [(a)] One agent is chosen randomly.
\item[(b1)] If the chosen agent is in state S, its state changes to I with probability $\lambda n(\mathrm{I}) dt$.
\item[(b2)] If the chosen agent is in state I, its state changes to R with probability $\mu dt$.
\item[(b3)] If the chosen agent is in state R, its state does not change.
\item[(c)] Processes (a) and (b) are repeated $N$ times.
\end{itemize}
The processes ((a), (b), and (c)) are repeated until there are no agents in state I in the network.

In this simulation, the number of infectious agents at the initial stage is set to be 10 and the following parameter values are used:
$\lambda = 0.2$, $\mu = 1$ and $dt = 0.01$.

\subsubsection{Spatial prisoner's dilemma}
The spatial prisoner's dilemma is a model for decision making with respect to a social dilemma. Each agent takes one of two strategies: C (cooperation) or D (defection). These two strategies are regarded as the states of the agents.

First, each agent occupies a node of the generated network and has an equal probability of choosing cooperation or defection as an initial strategy. All agents simultaneously update their strategy as follows:
\begin{itemize}
\item [(a)] Each agent plays the prisoner's dilemma game with all neighboring agents and receives the resulting payoff shown in Table \ref{table:payoff} ($T$ stands for the temptation of defection).
\item [(b)] Each agent imitates the strategy of the wealthiest among its neighbors. If an agent has the highest payoff among the neighbors, it retains its own strategy for the next iteration.
\end{itemize}
In this simulation, $T=1.5$, and $\epsilon= 0$, as in the work of \cite{nowak1992evolutionary}.

\begin{table}
\caption{Payoff matrix of the spatial prisoner's dilemma}
\label{table:payoff}
\begin{center}
\begin{tabular}{c|cc}
\hline
 &Cooperator & Defector \\
\hline\hline
Cooperator &1,1 & 0, $T$ \\
\hline
Defector &$T$,0 & $\epsilon$, $\epsilon$\\
\hline
\end{tabular}
\end{center}
\end{table}

\subsection{Node Categorizations}
\begin{figure*}[t]
\begin{center}
\includegraphics[width=0.9\linewidth]{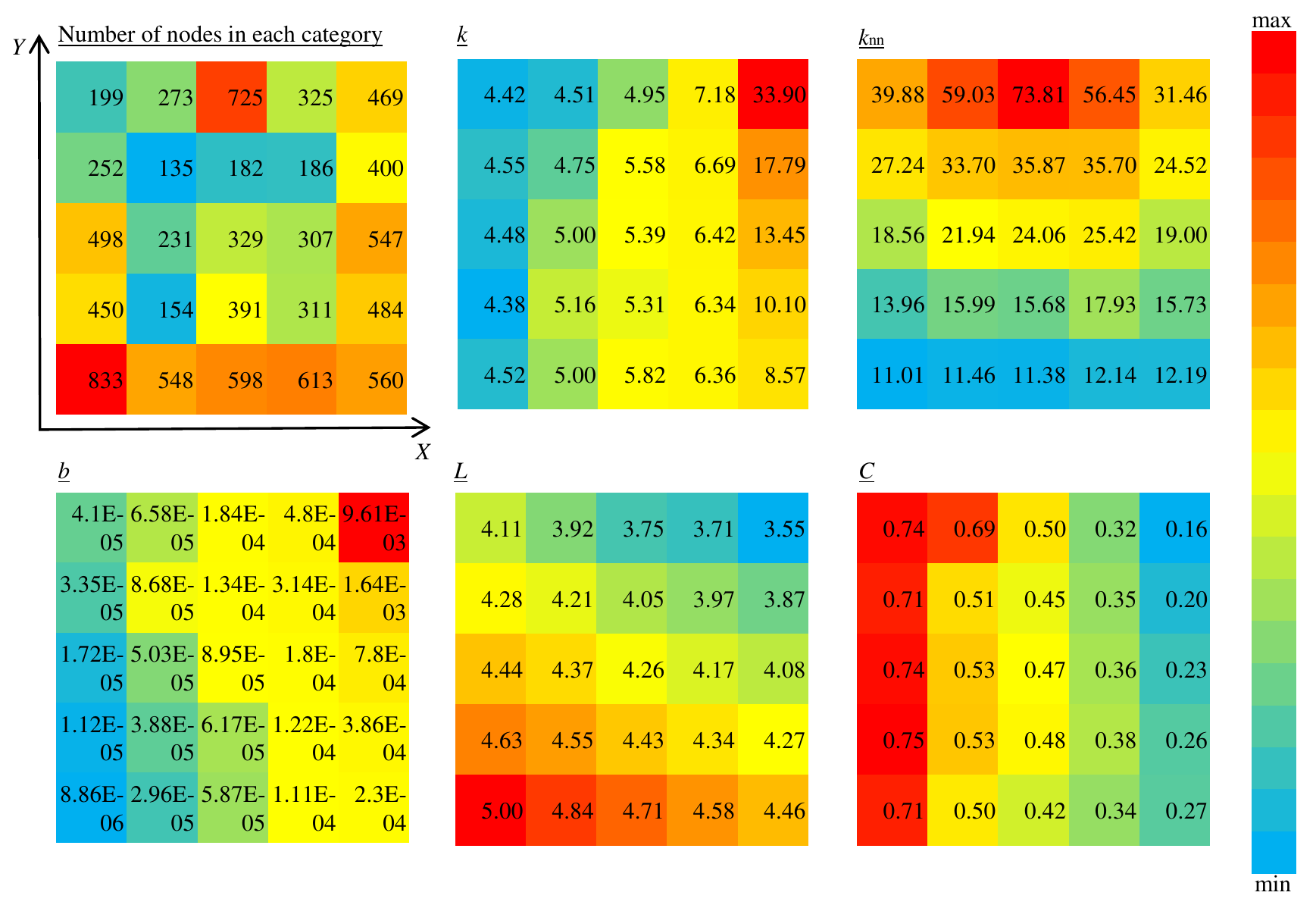}
\end{center}
\caption{Heat maps for the HK model}
\label{fig:HK_heat}
\end{figure*}
\begin{figure*}[t]
\begin{center}
\includegraphics[width=0.9\linewidth]{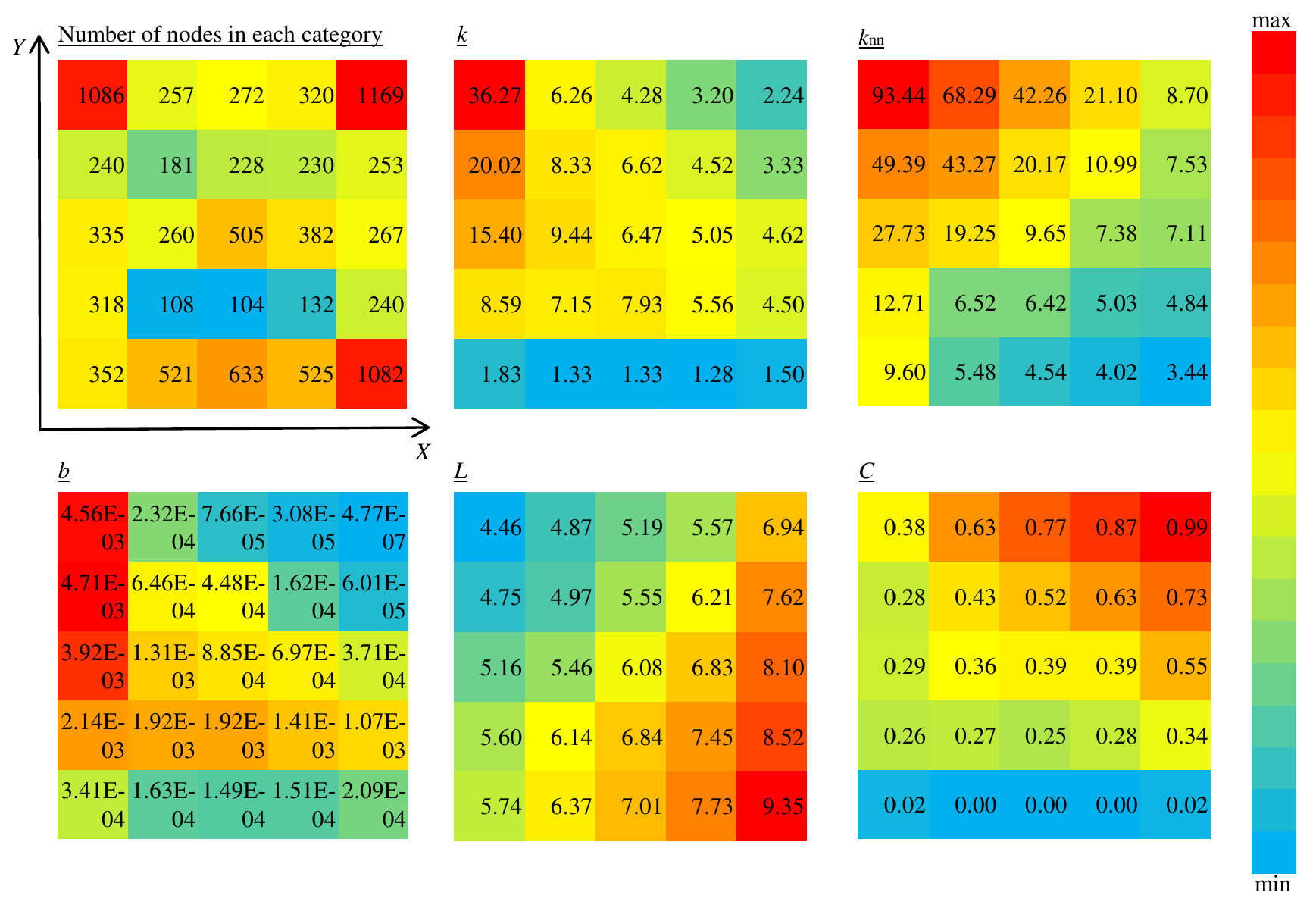}
\end{center}
\caption{Heat maps for the CNN model}
\label{fig:CNN_heat}
\end{figure*}
First, the property ${\bf n}_i = (k_i, k^i_{nn}, b_i, L_i, C_i)$ of the node $v_i$ is calculated for each node of the networks generated by the HK model and CNN model. Then, the nodes are categorized into the 5 by 5 lattice shown in Figure \ref{fig:lattice} using the SOM algorithm. 

Next, the heat maps for each component of the property ${\bf n}$ are created. The heat maps for the networks generated by the HK and CNN models are shown in Figures \ref{fig:HK_heat} and \ref{fig:CNN_heat}, respectively. In these figures, the number of nodes in each category is shown in the upper left panel of each figure.

For the network generated by HK model, each category is characterized as follows:
\begin{description}
\item[-] the degree $k$ increases for increasing $X$. The maximum value appears at $(4,4)$.
\item[-] the average degree of neighboring nodes $k_{nn}$ increases for increasing $Y$, and, for fixed $Y$, the maximum values is in the central $X$ region.
\item[-] the node betweenness $b$ increases for both increasing $X$ and increasing $Y$. The maximum value is at $(4,4)$ and is significantly larger than the values for other categories.
\item[-] the average path length $L$ decreases for both increasing $X$ and increasing $Y$. Compared to $b$, $L$ varies little from category to category.
\item[-] the clustering coefficient $C$ decreases for increasing $X$.
\end{description}

For the CNN model, each category is characterized as follows:
\begin{description}
\item[-] $k$ decreases for increasing $X$.
\item[-] $k_{nn}$ decreases (increases) for increasing $X$ ($Y$).
\item[-] $b$ decreases for increasing $X$.
\item[-] $L$ increases (decreases) for increasing $X$ ($Y$).
\item[-] $C$ increases for both increasing $X$ and increasing $Y$.
\end{description}

\subsection{Simulation of Epidemic Propagation}
\begin{figure}[t]
\begin{center}
\includegraphics[width=0.90\linewidth]{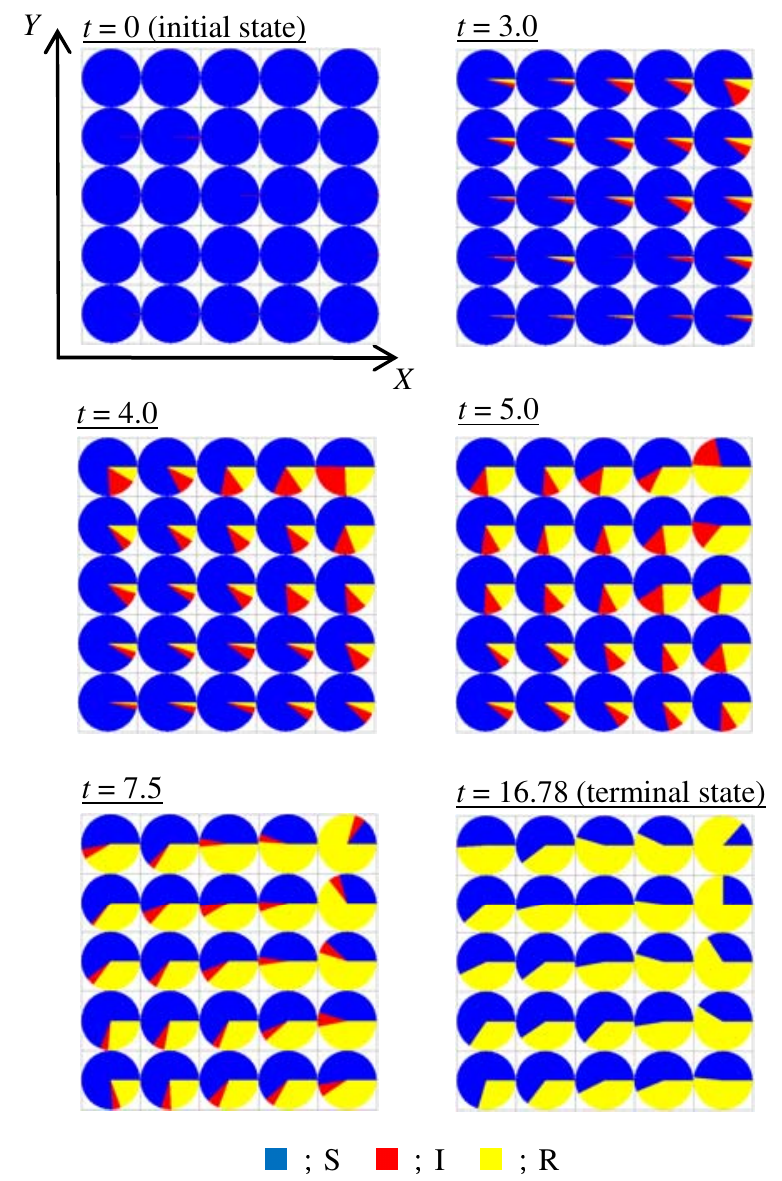}
\end{center}
\caption{SIR states over time for the network generated using the HK model}
\label{fig:HK_SIR}
\end{figure}

\begin{figure}[t]
\begin{center}
\includegraphics[width=0.81\linewidth]{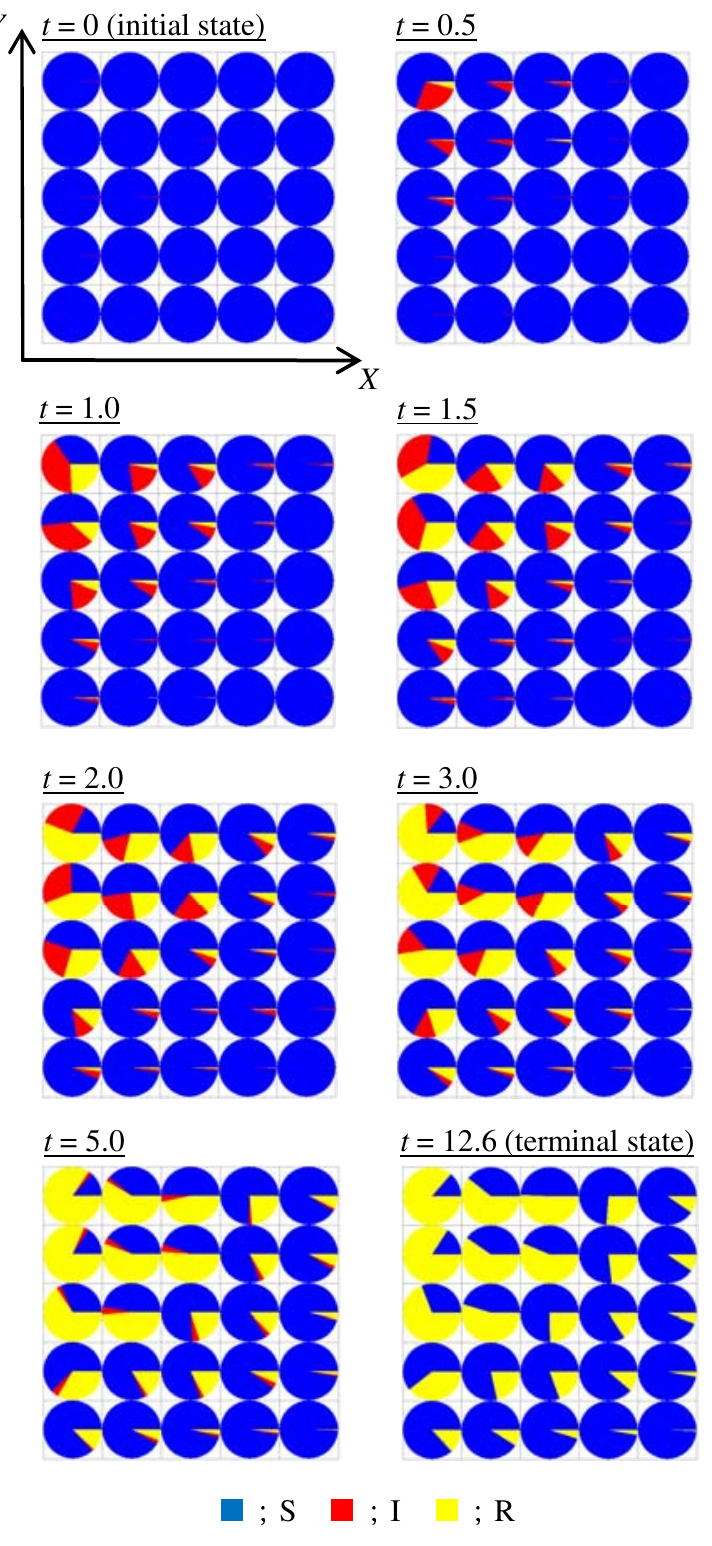}
\end{center}
\caption{SIR states over time for the network generated using the CNN model}
\label{fig:CNN_SIR}
\end{figure}

SIR states over time for the networks generated using the HK and CNN models are shown in Figures \ref{fig:HK_SIR} and \ref{fig:CNN_SIR}, respectively.

For the HK model, Figure \ref{fig:HK_SIR} shows that the initial outbreaks start around $(4,4)$ at $t=3.0$. Since $k$ is also largest in this category (Figure \ref{fig:HK_heat}), the model indicates epidemic outbreaks start near hubs, after which the epidemic expands. From $t=3.0$ to $4.0$, the epidemic spreads from  category$(3,4)$ to $(0,4)$ in the negative $X$ direction and from category$(4,3)$ to $(4,1)$ in the negative $Y$ direction. Referring to Figure \ref{fig:HK_heat}, these categories coincide with those which have large $k$ or $k_{nn}$, and short $L$. 
After $t=4$, the epidemic seems to spread toward category $(0,0)$. This means that the speed of propagation of the epidemic within categories with smaller $k$ and $k_{nn}$ and longer $L$ is relatively slow. At the terminal state ($t=16.78$), the pattern of the distribution of proportions of infected agents, which is shown by the distribution of state R, is almost the same as that of $b$ in Figure \ref{fig:HK_heat}. From this, one can conclude that agents on the shortest paths tend to transmit the epidemic.

For the CNN model, Figure \ref{fig:CNN_SIR} shows that a violent outbreak occurs in the category $(0,4)$ at $t=0.5$. As shown in Figure \ref{fig:CNN_heat}, this category has the largest $k$ and $k_{nn}$. Therefore, it is assumed that the mutual infections of agents in hubs causes a violent outbreak.
After $t=0.5$, the epidemic spreads from category $(0,4)$ toward $(4,0)$.
The proportions of infected agents are quite small in categories near $(4,0)$. As shown in Figure \ref{fig:CNN_heat}, these categories have small $k$ and $k_{nn}$ and long $L$. This means that the agents which have fewer links with and are distant from other agents have a lower probability of infection. 
This trend in the CNN model (Figure \ref{fig:CNN_SIR}) is more obvious than that in the HK model(Figure \ref{fig:HK_SIR}).
We also found that the pattern of the distribution of the proportions of infected agents at the terminal state ($t=12.6$) is almost same as those of $k_{nn}$ and $L$ shown in Figure \ref{fig:CNN_heat}, whereas, for the HK model, the pattern was similar to that of $b$.

\subsection{Simulation of Spatial Prisoner's Dilemma}
\begin{figure}[h]
\begin{center}
\includegraphics[width=0.85\linewidth]{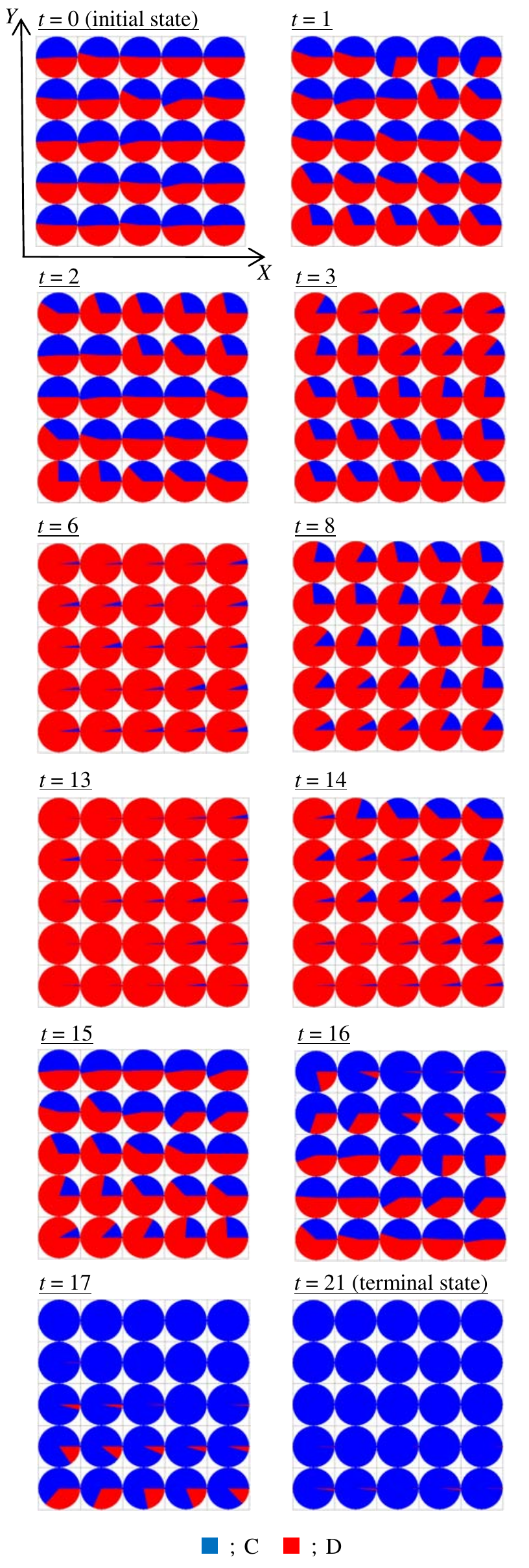}
\end{center}
\caption{CD states over time for the network generated using the HK model}
\label{fig:HK_SPD}
\end{figure}

The variation over time of the distribution of cooperators and defectors within the network generated by the HK and CNN models were visualized using the propose method. The visualized results for the HK model is shown in Figure \ref{fig:HK_SPD}. For simplicity, we show only the case in which cooperative agents are dominant.

For the HK model, cooperators have increased in several categories by $t=1$, in particular, in categories $(2,4)$, $(3,4)$, and $(4,4)$.
Referring to Figure \ref{fig:HK_heat}, the categories $(2,4)$ and $(4,4)$ are characterized by having the largest $k_{nn}$ and $k$, respectively. 
From $t=2$ to $6$, the defectors come to dominate the network. However, at $t=8$ the cooperators increase once more in categories $(2,4)$, $(3,4)$, and $(4,4)$. Later, the defectors again come to dominate, as shown at $t=13$, but then the cooperators increase again.
At this stage, the increase in cooperators starts first in categories characterized by large $k_{nn}$. The speed of the increase is then faster in categories with relatively large $k$.
Finally, the cooperators come to dominate almost the whole network, as shown at $t=21$. At this time, defectors survive only in categories with small $k_{nn}$, such as $(0,0)$ and $(0,1)$. 

For the CNN model, the variation over time of the distribution of cooperators and defectors as observed using the proposed method is similar to that for the network generated by the HK model. Defectors dominate initially, and cooperators expand outward from categories with large $k_{nn}$, starting from categories with large $k$. However, compared the results for the HK model, more defectors remain in categories with small $k$ or large $C$.

\section{Conclusions}
In this paper, we proposed a new analysis method for phenomena on networks based on a categorization of nodes. First, local statistics such as the average path length and the clustering coefficient for a node are calculated and assigned to the respective node to be used as the property of the node, denoted by multivariate variable ${\bf n}$.
Then, the nodes are categorized by applying the SOM algorithm to sets of ${\bf n}$.
Characteristic properties of some phenomena are visualized for each category.
The results are easily displayed in a two-dimensional lattice composed of the categories, even for the large-scale networks that existing visualization techniques cannot work well.
In our approach, the relationships between the phenomena and the network structure are revealed by the transition of the states of agents among the categories.

An epidemic propagation and a spatial prisoner's dilemma were examined using our method. In our analysis of two simulations, it was shown that the hubs play important roles on transmitting the state. In the case of the epidemic propagation, we found that the epidemic outbreak starts near hubs and continues by expand outward. In the spatial prisoner's dilemma, the visualization showed that cooperators expand outward from the cooperative hubs after the defectors have come to be dominant in almost the whole rest of the network. Although these results have been reported in previous studies \citep{moreno2002epidemic,tsukamoto2010spd}, several new findings, such as that the agents in categories with large $k_{nn}$ also promote the expansion of epidemic and cooperation, independent of the degree $k$, and that, for epidemic propagation, the pattern of the distribution of proportions of infected agents R is almost the same as that of node betweenness $b$ were both obtained by using the proposed method.
In future work, we will apply our method to the other kinds of simulations using networks, and then show the criteria which must be satisfied before our method may be applied.

%
%

\begin{acknowledgements}
This work was partially supported by Grant-in-Aid for Scientific Research (B) (21300031).
\end{acknowledgements}

\bibliographystyle{spbasic}      
\bibliography{reference_snam201104.bib}   

%
%

\end{document}